# A New Vehicle Localization Scheme based on Combined Optical Camera Communication and Photogrammetry

Md. Tanvir Hossan, Mostafa Zaman Chowdhury, Moh. Khalid Hasan, Md. Shahjalal, Trang Nguyen, Nam Tuan Le, and Yeong Min Jang
Email: mthossan@ieee.org

**Abstract—** The demand for autonomous vehicles is increasing gradually owing to their enormous potential benefits. However, several challenges, such as vehicle localization, are involved in the development of autonomous vehicles. A simple and secure algorithm for vehicle positioning is proposed herein without massively modifying the existing transportation infrastructure. For vehicle localization, vehicles on the road are classified into two categories: host vehicles (HVs) are the ones used to estimate other vehicles' positions and forwarding vehicles (FVs) are the ones that move in front of the HVs. The FV transmits modulated data from the tail (or back) light, and the camera of the HV receives that signal using optical camera communication (OCC). In addition, the streetlight (SL) data is considered to ensure the position accuracy of the HV. Determining the HV position minimizes the relative position variation between the HV and FV. Using photogrammetry, the distance between FV or SL and the camera of the HV is calculated by measuring the occupied image area on the image sensor. Comparing the change in distance between HV and SLs with the change in distance between HV and FV, the positions of FVs are determined. The performance of the proposed technique is analyzed, and the results indicate a significant improvement in performance. The experimental distance measurement validated the feasibility of the proposed scheme.

## 1. Introduction

Localization refers to the process of identifying the location (x and y coordinates in two-dimensional (2D) space and x, y, and z coordinates in three-dimensional (3D) space) of an object in a certain point in space at a specific time. Several studies are contributing to the development of accurate localization schemes owing to increased demand for Internet-of-Things (IoT) applications. The necessity of a localization scheme is integrated within the requirement of IoT. IoT relies on an enormous number of physical objects (e.g., sensor nodes and sensor networks) that are connected via the Internet [1]. These objects can be interconnected to each other either via wire or wireless mediums. A localization scheme is an important concern for connecting sensor nodes in remote location. A node cannot access or wirelessly communicate with other nodes without accurately positioning itself. The characteristics of localization schemes vary with the features of indoor and outdoor environments [2].

It is well known that localizing sensor nodes indoors can be a crucial obligation for modern businesses and commerce. However, issues related to outdoor localization, particularly vehicle localization, are prioritized over indoor localization. Recently, following road traffic safety [3] has become important owing to the increasing number of fatal road accidents. World Health Organization statistics [4] shows that traffic-related accidents worldwide resulted in 1.3 million deaths of people between 15 and 29 years, and the number of non-lethal injuries is 15–40 times greater (between 20 and 50 million). Thus, traffic fatalities rank among the 10 top causes of death, comparable to suicide, HIV/AIDS, homicide, and other diseases. The most common cause of traffic fatalities (around 60%) is high vehicle speeds (above 80 km/h) on the road [5]. Autonomous vehicles can help minimize traffic deaths. Meanwhile, the demand for autonomous vehicles has been rising dramatically to avoid accidents [6]. Furthermore, outdoor localization is of prime importance in the transportation domain, particularly, for autonomous vehicles, which requires localizing other vehicles from the host vehicle (HV) in road environments such as highways. For autonomous vehicles, the features of localization are classified as active and passive. Active features include setting regions of interest (ROI) and measuring the possibility of communicating with other vehicles and maintaining safe distance from other vehicles to avoid unwanted collisions by measuring spatial and temporal scenarios [7]. Passive features include obtaining localization information from individual vehicles, which can then be accumulated by a traffic control center and utilizing in effective way to mitigate traffic congestion.

*1.1 Existing solutions, limitations, and current trends in vehicle localization.* Global positioning system (GPS) is considered as the most prominent solution for outdoor localization scheme. GPS provides a line-of-sight vehicle localization solution using the sensor information from the [8]–[10] and data from a satellite orbiting at an altitude of approximately 20,000 km. GPS uses the radio frequency (RF) band for positioning the HV on road. However, the HV cannot measure its own distance from other vehicle, such as the forwarding vehicle (FV), via GPS; it offers only the current location of the HV. Moreover, this localization scheme is fraught with several challenges, such as GPS signals being blocked by obstacles such as buildings, subway, tunnels, and trees. Localization using GPS can generate a localization error of up to 1 m within 10 s [11]. A wireless network standard for vehicle states, called IEEE 802.11p [12], [13], is available; this is referred to as wireless access in vehicular environments (WAVE) [14]. This standard is used to maintain a communication network among vehicles within vehicular ad hoc networks (VANETs) [15] and to support intelligent transport system applications. RF signals in VANET systems are used for communication and vehicle localization [16]. Owing to various environmental effects and the multipath nature of the network, non-Gaussian noise is included with the transmitted signal, whose strength shows nonlinear characteristics over distance. The WAVE standard uses a license-free RF band (i.e., 2.4 GHz) [17], which is open to interference from other signal sources, thus making the entire network vulnerable from the viewpoints of both communication and localization. Other existing technologies for vehicle localization include light detection and ranging (LiDAR) [18]–[20] and the time-of-flight (ToF) camera technique [21]–[24]. Light-emitting diodes (LED) and cameras or photodiodes are embedded in LiDAR and ToF system infrastructures; however, they are used only for detection and ranging. These equipment are not useful for vehicle-to-vehicle or vehicle-to-infrastructure communications [25]–[29] and are expensive to be used in a vehicular environment.

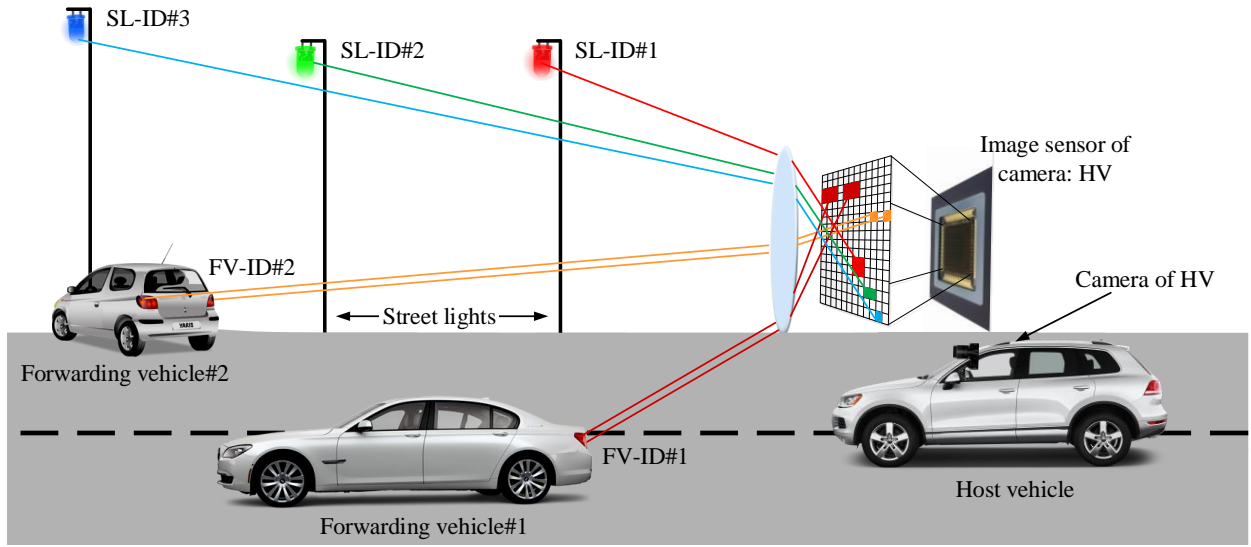

FIGURE 1: OCC and photogrammetry-based vehicles localizing by comparing relative position with the help of streetlights.

Optical wireless communication (OWC) is an emerging and promising technology [30] that is viable for handling scenarios wherein RF faces challenges. OWC is not intended to replace RF; however, the coexistence of both can provide a better solution [31] for communication and localization. Optical camera communication (OCC) [32] is a sub-area of OWC that uses a camera as a receiver to decode signals from a modulated light source, e.g., LEDs, by varying the state of the light source to transmit binary data via optical channels. It is a secure, safe, reliable, and fast method for communication as well as localization [33]. A unique feature of OCC is that the camera used for vehicle localization can simultaneously be used to communicate with other vehicles that transmit signals using modulated lights. With little modifications, LEDs in existing infrastructure i.e., vehicles and streetlights (SLs) can be used for communication (e.g., bidirectional communication between two vehicles or between vehicles and infrastructure) [34]–[40].

To better communicate in outdoor environments, vehicles around the HV must be localized precisely. More importantly, multiple-input and multiple-output (MIMO) features of OCC [41] should allow the HV to simultaneously communicate with more than one vehicle. In [42], author presents a received signal strength based visible light communication localization scheme, but it could not improve localization performance such as more complex models of the environment or additional hardware are required for localization. The localization of multiple vehicles would require incorporating OCC and photogrammetry technologies [43]. Photogrammetry [44], [45] deals with a branch of geometry wherein an image sensor (IS) is used to measure an object by quantifying the photon intensities of different wavelengths of light incident on an area, i.e., a unit pixel of a camera. Photogrammetry helps accumulate information on semantic and geometric properties, and variation of relative distances of objects, which refers to vehicles in this context. This vehicle location information can be shared with following vehicles with the help of OCC and rear-facing LED lights. Figure 1 shows a vehicle localization scheme combining OCC and photogrammetry.

A vehicle localization technique, wherein each FV broadcasts its identity (ID) to the HV as FV-ID, is proposed herein. After extracting the unique ID from the received signal, the HV can distinguish an FV from other FVs. Since the HV and FV simultaneously change their positions over time, location of the HV should be normalized based on the location of a fixed object, for instance, a SL. Comparing the locations of more than one SL relative to the HV, a virtual location of the HV can be temporarily generated. This HV location information acts as the origin of a Cartesian coordinate system that allows determining the FV location relative to the HV.

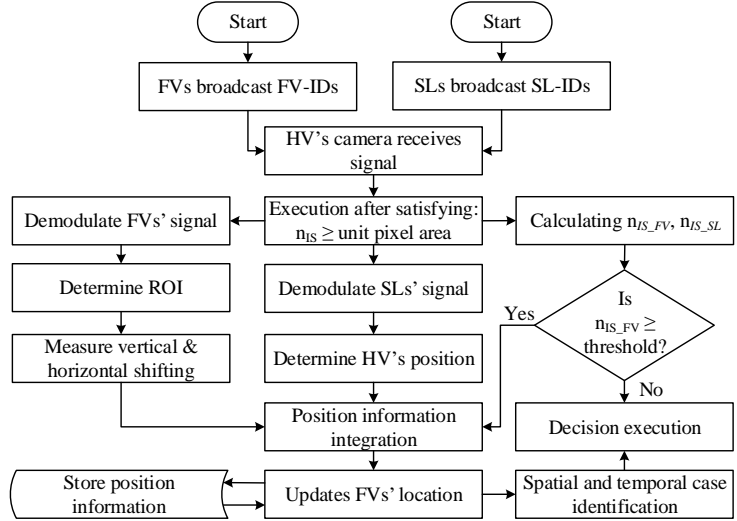

FIGURE 2: Flow chart for a vehicle localization technique based on OCC and photogrammetry.

For autonomous vehicle localization, infrared LED array can be attached at the SLs and back side of the FVs to clarify the area of the LED array. Though the near-infrared (NIR) source is visible to the camera, it is possible to receive data from the FVs and SLs. Detecting light intensity for the nearest light sources is much higher on the IS of the camera of HV rather than far distance light source. Compared with visible-light-based communication, NIR-based communication is influenced by the optical channel. Under daylight, it is a challenging to receive data from an NIR-based transmitter. Importantly, a recent development of high-dynamic range imaging technique reduce noise and enhance the image quality under daylight [46], [47]. Therefore, it is expected that ambient light no longer poses a problem for OCC, even when the transmitter possesses an NIR-optical band. Simulation results show vehicle localization accuracy with considering the impact of several parameters including signal-to-interference-plus-noise ratio (SINR), IS resolution, camera exposer time, and the distance between two SLs.

The remainder of this paper is organized as follows: Section 2 explains a detailed theoretical and mathematical model of our proposed scheme. Experimental set-up for distance measurement is shown in Section 3. In Section 4 the simulation results associated with vehicle localization studies is presented. Finally, Section 5 presents a summary of lessons learned and concludes this study.

## 2. Development of Proposed Scheme

Almost every vehicle produced in recent years is equipped with a camera (i.e., less than 30 frames/sec) that is used to monitor the outdoor scenarios and to assist the drivers by providing a view of their blind spot. Herein, the HV communicates with the FV and measures the distance between vehicles using such a camera mounted in front of the vehicle. Using OCC, this camera detects transmitted signal IDs, such as FV-ID and SL-ID from each FV and SL simultaneously. A pair of taillight on FVs transmits ID in different phases to modulate the data using a modulation scheme called spatial-two-phase-shift-keying (S2-PSK) [48] to the HV's camera. These LEDs transmit at a constant clock rate (e.g., 125 or 200 Hz) to send a flicker-free signal. The SLs use the same modulation scheme as the FV for transmitting the SL-ID. MIMO is a distinctive functionality of a camera that helps distinguishing FV-ID from SL-ID. These IDs are required to determine the ROI for vehicle localization. The ROI specifies the camera's viewing region within an image and helps minimize the scope of false-position results from the main event. On the road, an FV can move side-to-side or change its direct distance with respect to the HV, which we are stated as horizontal shift and vertical shift, respectively. These position shifts lead to a change in image size that can be measured from the IS. Both the FV and HV move simultaneously; therefore, it is not always possible to localize the position of the FV relative to the HV. However, if the position of the HV is known, the relative positions of the FV and HV can be easily compared. The position of SL is fixed relative to every vehicle on the road; therefore, it is necessary to receive SL-IDs from the SLs to determine the HV position. Figure 2 shows a flowchart of the proposed localization scheme wherein the FV location information is compared with the current HV location to identify special and temporal cases. After receiving IDs, algorithm will move ahead if the size of detecting image area of FV is greater or equal to unit pixel area. In decision symbol of the algorithm, the threshold value indicates the minimum distance between HV and FVs to avoid collision.

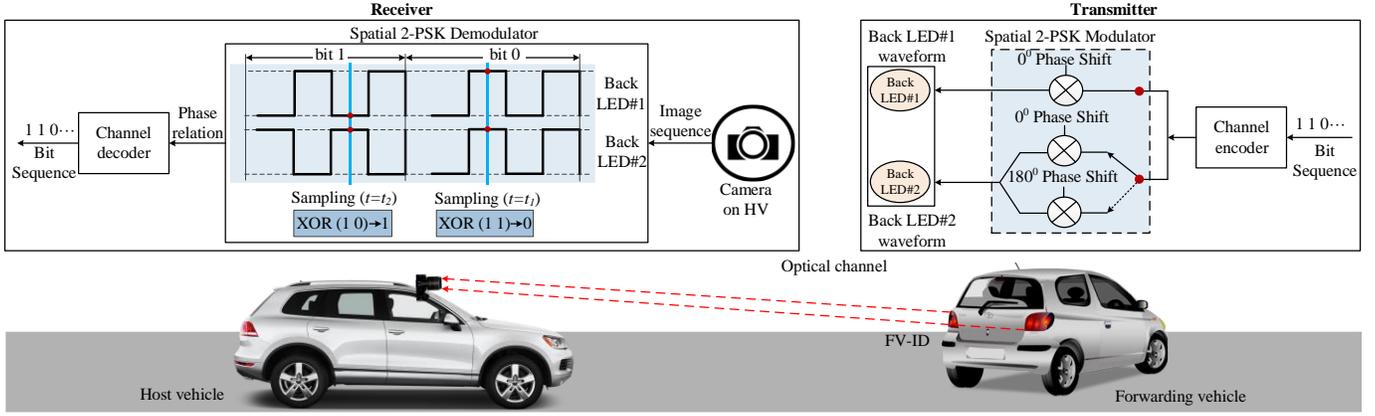

FIGURE 3: Coding and decoding LED-ID using OCC.

*2.1. LED-ID from SL and FV.* In Figure 3, the two LED pairs fixed on the back of the FV transmit a modulated FV-ID [48]. The SL transmits the SL-ID using the same modulated signal (i.e., S2-PSK) by dividing a single LED array into two pairs of LED arrays. Depending on the input bit sequence, the transmitting signal phases of the LED array pairs can differ. The scheme uses a symmetric Manchester symbol to map each LED symbol. Using a spatial under-sampling approach, the LED pairs transmit in the same phase for bit 0 and in different phases for bit 1. The bit interval $s_1(t)$ for one of tail LED is as follows

$$s_1(t) = \sum_{k=0}^{N} s_1(t_k + kT) \quad \text{where} \begin{cases} 0 \leq t < T_{bit} \\ s_1(t_k) = \begin{cases} 1, & 0 \leq t_k < T/2 \\ 0, & T/2 \leq t_k < 0 \end{cases} \end{cases} \quad (1)$$

where $k$ is an unsigned integer of $N$ the bit-interval cycles, $T_{bit}$ is a bit interval, and $T$ is the cyclic interval of signal.

The bit interval $s_2(t)$ for other of tail LED is as follows

$$s_2(t) = \sum_{k=0}^{N} s_2(t_k + kT) \quad \text{where} \begin{cases} 0 \leq t < T_{bit} \\ s_2(t_k) = \begin{cases} 1, & \overline{s_1(t_k)} \\ 0, & s_1(t_k) \end{cases} \end{cases} \quad (2)$$

From the same camera image, the S2-PSK demodulates a bit from a pair of states of two different LEDs. At sampling time $t_s$, the same states of two tail LEDs on the same image resemble bit 0, otherwise bit is 1. XOR operation determines the value of bit captured in the same image as follows

$$bit = s_1(t_s) \oplus s_2(t_s) \quad (3)$$

where $s_1(t_s)$ and $s_2(t_s)$ are states of two LEDs at sampling time $t_s$.

Compared with other modulation schemes (e.g., undersampled phase shift on-off keying) [49], this demodulation can gain a lower bit error rate (BER) within an image. A nonlinear XOR classifier can remove the remaining BER. The BER performance of this modulation scheme [41] is stated as follows

$$P_{e,S2-PSK} = 2\alpha p_e (1 - \alpha p_e) \quad (4)$$

where $p_e$ is the bit error probability of LED state and $\alpha$ is the error rate enhancement.

Considering environmental effects, the SINR [50] is expressed as follows

$$SINR = \frac{(\kappa P_{opt} H)^2}{t^2 N_0 B + \sum (\kappa P_{opt} H_{else})^2} \quad (5)$$

where $\kappa$ is the optical-to-electric conversion efficiency at the camera, $N_0$ is the noise power spectral density, $B$ is the modulation bandwidth, $H$ is the optical channel gain, $H_{else}$ channel gain for interfering light sources, $P_{opt}$ average optical power, and $t$ is the conversion between average electrical power $P_{elec}$ and $P_{opt}$.

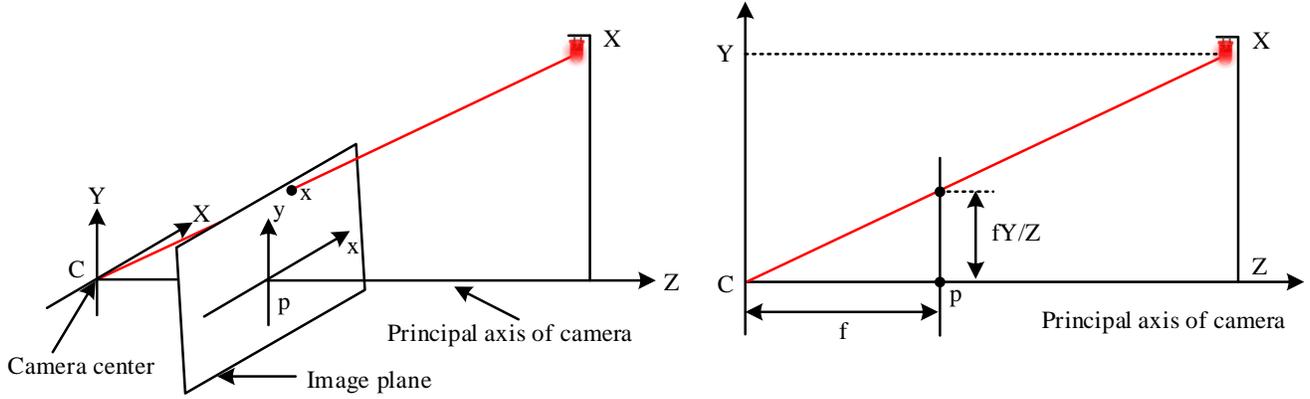

FIGURE 4: Camera calibration for vehicle localization.

Meanwhile, the optical channel gain is expressed as follows

$$H = \left\{ \frac{(m+1)A_c}{2\pi D^2} g(\theta)T_s(\theta)\cos^m(\phi)\cos(\theta) \right. \tag{6}$$

where $m$ is the Lambertian index, $A_c$ is the physical area of IS, $D$ is the distance between transmitter and receiver, $\theta$ is the angle of incidence, and $\phi$ is the angle of irradiation.

*2.2. Camera calliberation and photogrammetry.* In computer vision applications, camera calibration is essential to determine real-world coordinates from simple 2D images. The simplest camera calibration method involves using a pinhole camera model to provide a perfect perspective transformation [51]. In a Euclidean coordinate system, the origin of the projected object coordinates is shifted from the principal point of the camera's image plane, as shown in Figure 4. Mapping an object's Euclidean three-space $\mathbb{R}^3$ coordinates $(X,Y,Z)^T$ to Euclidean two-space $\mathbb{R}^2$ allows for the mapping of an object from the real world to image coordinates as follows

$$(X,Y,Z)^T \rightarrow \left( \frac{FX}{Z} + p_x, \frac{FY}{Z} + p_y \right)^T \tag{7}$$

where $F$ is the focal length of the camera and $(p_x, p_y)^T$ are the principal point coordinates of the camera.

Homogeneous vectors allow us to map the coordinates of the real-world and an image in terms of matrix multiplication as follows

$$\begin{pmatrix} fX + Zp_x \\ fY + Zp_y \\ Z \end{pmatrix} = \begin{bmatrix} R & -RC \\ 0 & 1 \end{bmatrix} \begin{bmatrix} F_x & s & p_x & 0 \\ & F_y & p_y & 0 \\ & & 1 & 0 \end{bmatrix} \begin{pmatrix} X \\ Y \\ Z \\ 1 \end{pmatrix} \tag{8}$$

where $F_x (= Fm_x)$ and $F_y (= Fm_y)$ represent the focal length of camera in terms of pixel area along the $x$ and $y$ direction, respectively; $s$ is the skew parameter, and it is normally zero; $R$ is the camera's orientation relative to real-world coordinates; and $C$ denotes the camera's coordinates. Here, $m_x$ and $m_y$ denote the number of pixels per unit distance, expressed as image coordinates in the $x$ and $y$ direction, respectively. Equation (8) can be expressed succinctly as follows

$$x = RK[I \mid -C]X \tag{9}$$

where $K$ is the calibration matrix of the camera, $X$ is the coordinate matrix in a world coordinate frame, and $I$ is an identity matrix.

Let the distance from the LED to the camera lens be $D$, and the distance from the focal point of the camera to the projected image on the IS be $e$. Then, the ratio of LED distance and image distance is distance as follows

$$\frac{D}{e} = \frac{D}{F} - 1 \tag{10}$$

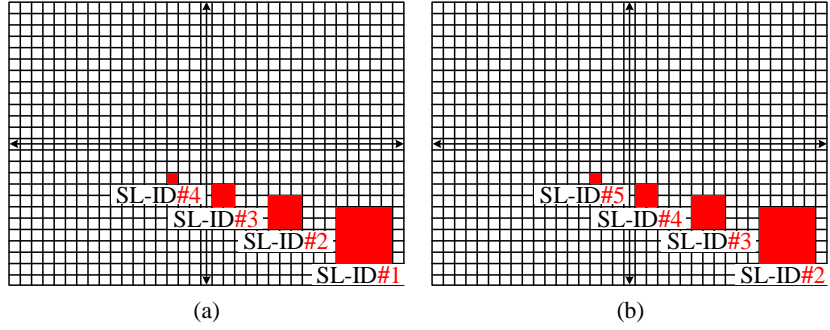

FIGURE 5: SL-IDs change with the change of HV's position (a) at time *t* to (b) at time *(t + 1)*.

Image area calculation performance depends on $F$ and $D$, and it must satisfy the condition $F \ll D$. Therefore, $(D-F)$ is equivalent to $D$. The ratio height and width of a LED $(l, r)$ and same ratio of the projectile image $(l_i, r_i)$ is set of the magnification of camera lens. This ratio is similar to the ratio of LED distance $D$ and image distance $e$, which is described as follows

$$l_i r_i = \frac{F^2 lr}{D^2} \qquad (11)$$

The number of pixel on IS for a particular object, $n_{IS}$ is the ratio of projected image area to the unit pixel area of the IS. In an IS, the unit pixel length is $\rho$, unit pixel area is $\rho^2$ and $A$ is the area of the LED light source. Thus, the following equation can be stated from (11) as follows

$$D = \frac{F}{\rho} \sqrt{\frac{A}{n_{IS}}} \qquad (12)$$

Distance is always an absolute value; therefore, the negative sign in (12) can be discarded. If the camera focal length $F$ and unit pixel length $\rho$ are maintained constant for a certain camera, the distance of the LED is kept proportional with respect to the square root of the LED's area and disproportional with respect to the square root of pixel area of that LED on the IS [39].

*2.3. Determining HV position.* The origin in coordinate systems, such as Cartesian and polar coordinate systems, is required to determine the position of one or more objects in either 2D or 3D space. In an outdoor environment, nearly every vehicle frequently changes its location; the measurement distances from HV to FV are not always accurate because of subsequent variations in their relative positions over time. Therefore, the location of FV from the origin or any stable location cannot be measured, and it is better if the HV location is known throughout this period. The shift in the FV's location can be measured by comparing is location with the current location of the HV.

In our proposed localization scheme measures the HV's location by comparing it with the location of SLs. SLs' location is always fixed with respect to other vehicles within this mobile scenario. This distance comparison yields location information for the HV, which also represents its virtual coordinates. To ensure the accuracy of this measurement system, location information from the on-board diagnostic II (OBD II) system and SLs is combined by the HV. The SL-ID should contain unique information that helps to distinguish this ID from other transmitted signals, such as FV-ID. The header of the SL-ID indicates that this ID belongs to a specific SL. In addition, other information, such as height of the SL from the ground and distance between two SLs on the same road, can be added after the header of the ID. At the same direction, there is a similarity among all SL-IDs of the SLs and a unique value within the IDs increasing or decreasing gradually.

After selecting the ROI, the distance between the camera and LED of the SL is measured using photogrammetry. Figure 5 shows the change in getting a SL-ID within the field of view (FOV) of the camera owing to a change in the HV position. The two axes are used to show the midpoint of IS. In Figure 5(a), the ID from the SL shows *SL-ID#1 ~ SL-ID#4* at time *t*. These IDs vary *from SL-ID#2 ~ SL-ID#5* at time *(t+1)*, which is shown in Figure 5(b). The size of the projectile image area of the nearest SL occupies a greater area on the IS compared with other SLs. The direct distance is calculated using Equation (12); the distance for *SL-ID#1* is shorter compared with that for *SL-ID#4*, as shown in Figure 5(a).

Using OCC, the camera decodes the SL-IDs of the SLs. Figure 6(a) shows, SL's height $(SL\_h_n)$, and the constant distance between two SLs is $d_n$ where $n^{th} (n = 1, 2, ..., \mathbb{N})$ related to the number of SL. Using photogrammetry, $D_{SL_j - HV}$ is determined as the

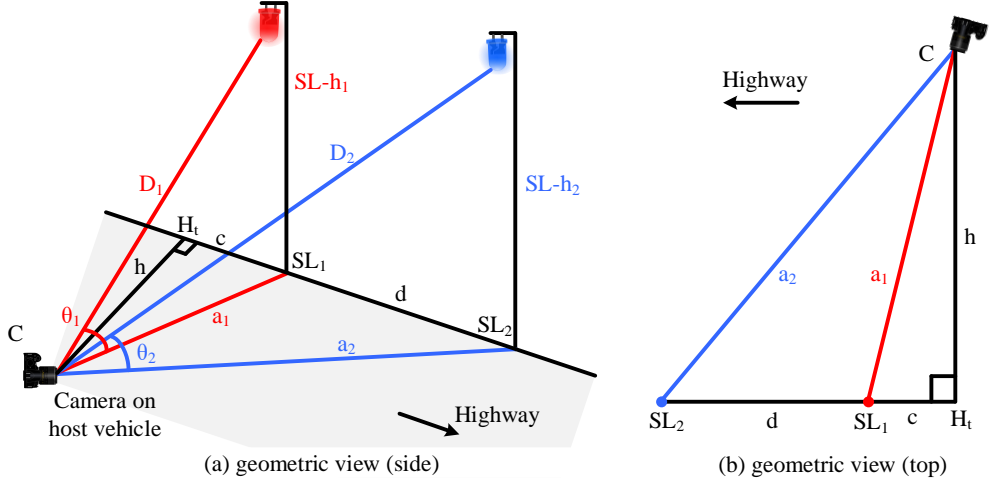

(a) geometric view (side)          (b) geometric view (top)

FIGURE 6: Obtaining virtual coordinates from (a) road scenario and (b) measuring vertical distance from the HV to pavement.

measured direct distance between the camera and SL's LED where $j^{th}(j = 1, 2, ..., \mathbb{N})$ states the number of iteration sequence over a period. The horizontal distance between the camera and SL is $a_j$. These horizontal distances are calculated by applying Pythagorean theorem on a right triangle where $D_{SL_j-HV}$ and SL's height are the remaining two sides of that right triangle.

From the top geometric view, a few triangulations can be generated after decoding this distance information. At a certain time, applying Pythagorean theorem again to triangles $CSL_1H_t$ and $CSL_1SL_2$, we obtain

$$a_1^2 = c^2 + h^2 \tag{13}$$

$$a_2^2 = (d+c)^2 + h^2 \tag{14}$$

where $h$ is the horizontal distance from the camera to the pavement, $c$ is the distance between the cross point of the horizontal line and the shortest distance from the cross point to the SL, $a_1$ is the horizontal distance for $SL_1$, and $a_2$ is for $SL_2$ shown in Figure 6(b). In all cases, $c < d$. We combine Equation (13) and (14) as

$$c = \frac{(a_2^2 - a_1^2) - d^2}{2d} \tag{15}$$

The horizontal distance is determined from the camera to the pavement by combining Equations (13) and (15) as follows

$$h = \sqrt{a_1^2 - \left\{\frac{(a_2^2 - a_1^2) - d^2}{2d}\right\}^2} \tag{16}$$

The position of HV is a function that varies with the horizontal position $h_j$, which is always positive; angular position $\theta_{SL_j-HV}$ of the SL relative to the HV; SL's LED image area is $n_{IS\_SL}$ on IS; and velocity of HV $V_{HV}$. When the HV moves, the parameters related to the HV's position change. If the initial position is recorded at time $t$; after $\Delta t$, the position of HV states as follows

$$P_{HV}(t + \Delta t) : \left\{ h_j \pm \Delta h; \; \Delta\theta_{SL_j-HV}; \; n_{IS\_SL} \geq \rho^2; \; V_{HV}\left(\Delta[c_j + d_n]/\Delta t\right) \right\} \tag{17}$$

The horizontal distance between the HV and the pavement is a function of the horizontal direct distance $a_j$, distance between two SLs i.e., $d_n$; and distance between the cross point of the horizontal line and the shortest distance from the cross point to the SL i.e., $c_j$, where all these values are also changed according to the change of the angular position $\theta_{SL_j-HV}$

$$h(a_j, \; c_j, \; d_n) : \left\{ \Delta\theta_{SL_j-HV} \right\} \tag{18}$$

The direct distance between the SL and HV depends on the area of the SL's LED $n_{IS\_SL}$ on the IS and the angular position $\theta_{SL_j-HV}$. If the value $n_{IS\_SL}$ is less than the unit pixel area $\rho^2$ ( this will happen when the position of the SL is too far from the HV); the value of image area is ignored from the calculation. On the hand, angular position $\theta_{SL_j-HV}$ changes with the bending of the road (or the edge of a road) and the $D_{SL_j-HV}$ changes accordingly. Therefore, the following expression is stated for measuring direct distance

$$D_{SL_j-HV}(n_{IS\_SL}, \Delta\theta_{SL_j-HV}) : \{n_{IS\_SL} \geq \rho^2; \Delta\theta_{SL_j-HV}\} \tag{19}$$

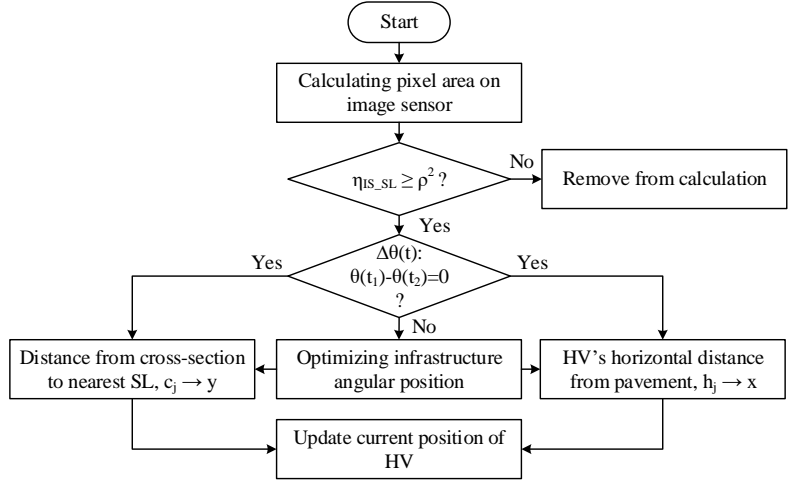

FIGURE 7: Flowchart of detailed development of the HV's location information.

From initial time $t$ to $\Delta t$, the changes of angular position is found by simply comparing current angular position with previously recorded value as follows

$$\Delta\theta_{SL-HV}(t+\Delta t) : \{\theta_{SL_1-HV} \sim \theta_{SL_2-HV}\} \tag{20}$$

The horizontal distance $h_j$ is set as the x-coordinate and $c_j$ is set as the y-coordinate for the HV with respect to the nearest SL. Therefore, when the HV moves, this distance information is updated. Figure 7 shows the flowchart for calculating and correcting horizontal position information. From eq. (20), a considerably initial angular position $\theta_{SL_1-HV}$ compared with the conjugate angular position $\theta_{SL_2-HV}$ indicates the presence of a ground curvature. In such cases, infrastructure effect optimization is required; otherwise, the horizontal distance can be easily updated.

*2.4. Derteming the position of FVs from the HV.* Each vehicle has a pair of headlight and taillight. Using OCC, taillights of FVs transmit modulate signals to a receiver, i.e., a camera of the following vehicle. Using this modulated signal, the FV transmits emergency information along with some basic vehicle information, e.g., the area of the single light from the rear of the vehicle. This transmitted signal from one pair of taillights is noted as FV-ID, and one ID is unique compared to other vehicles' IDs.

For a proper communication among vehicles (i.e., FVs and HV) on the road, the signal transmitted from both taillights must be received by the camera. There are scenarios in which signal interruption can occur, for instance, the HV may monitor two vehicles from an angle wherein one of the lights from a single vehicle is covered by the other vehicle. In this case, data extraction is not possible although a single light signal is received by the HV's camera. Moreover, two vehicles can be differentiated using their FV-IDs even if they are moving in parallel. The advantages of LED-ID based vehicle identification make it possible to fix the ROI, which is the preliminary condition for successful communication and localization. Figure 8 shows every FV broadcasting FV-IDs along with the SL as SL-IDs. In Figure 8(a), the background is turned black by controlling the shutter speed of the camera mounted in the HV. After demodulation and decoding of the transmitted signal, all IDs are accumulated, as shown in Figure 8(b).

The area of taillight LED arrays on the IS changes relative to changes in the distance between the HV and FV. By calculating the area of these images on the IS, two types of FV position shifts can be determined with respect to the HV, namely, horizontal and vertical shifting. Horizontal FV shift will be visible if the vehicle changes its position from side-to-side. Concurrently, the vehicle can slow down, changing the direct distance between the FV and HV, which is defined as vertical shift. The left of Figure 9 shows an image of the taillight LED that is projected on the left side of the image after being refracted by the camera lens. The image of the original light source is on the HV's right side. Furthermore, the middle and right parts of Figure 9 show that the vehicle is moving from the right to the middle and later to the left with respect to the HV. By contrast, in the left part of Figure 10, the area of the projected image is smaller than the other two parts, which shows that the FV was initially far from the HV and that this distance decreased gradually.

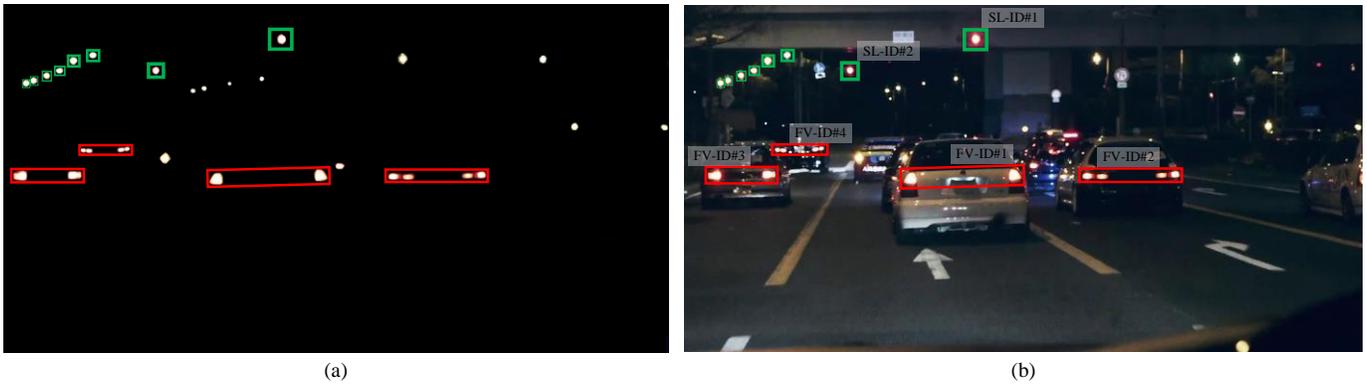

FIGURE 8: Using OCC, (a) selecting region-of-interest and (b) receiving IDs.

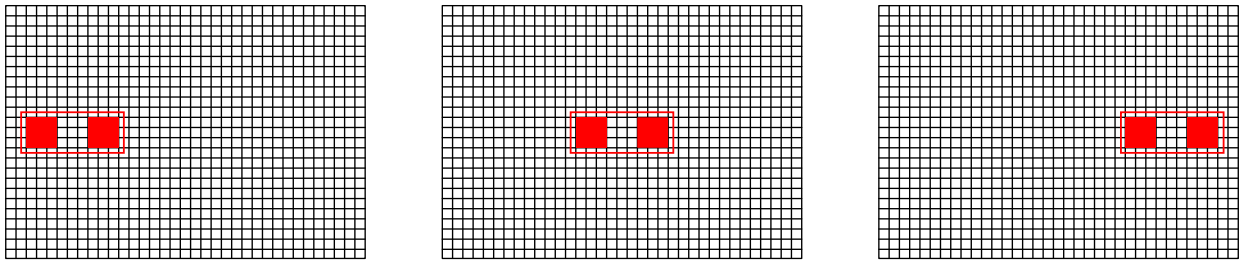

(i) FV at the right side of the HV    (ii) FV stays straight of the HV    (iii) FV at the left side of the HV

FIGURE 9: A pair of FV's taillights moves from the (i) left to the (ii) middle, and finally to the (iii) right on image sensor; implying that the FV is moving right to left with respect to the HV.

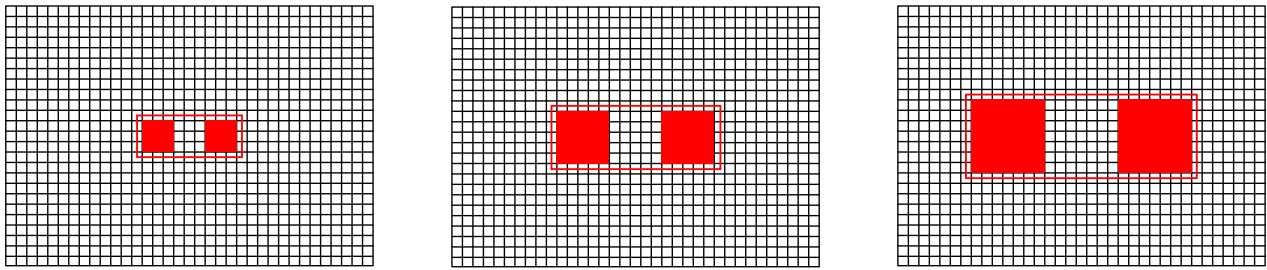

(i) FV keeps safe distance from HV    (ii) FV stays min. distance from HV    (iii) Critical distance between FV-HV

FIGURE 10: A pair of FV's taillight increases in size gradually from (i) left, (ii) middle, and (iii) right on image sensor; implying that the FV is moving closer to the HV.

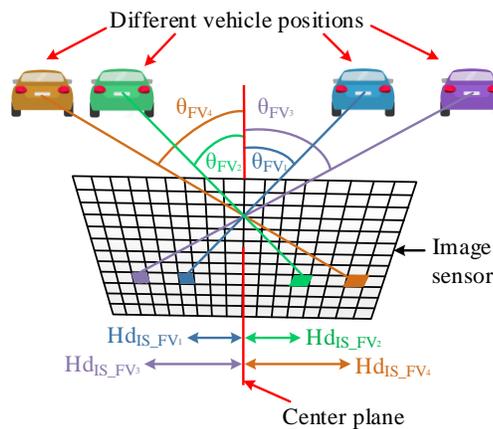

FIGURE 11: Measurement of vehicular angular position from horizontal displacement on image sensor.

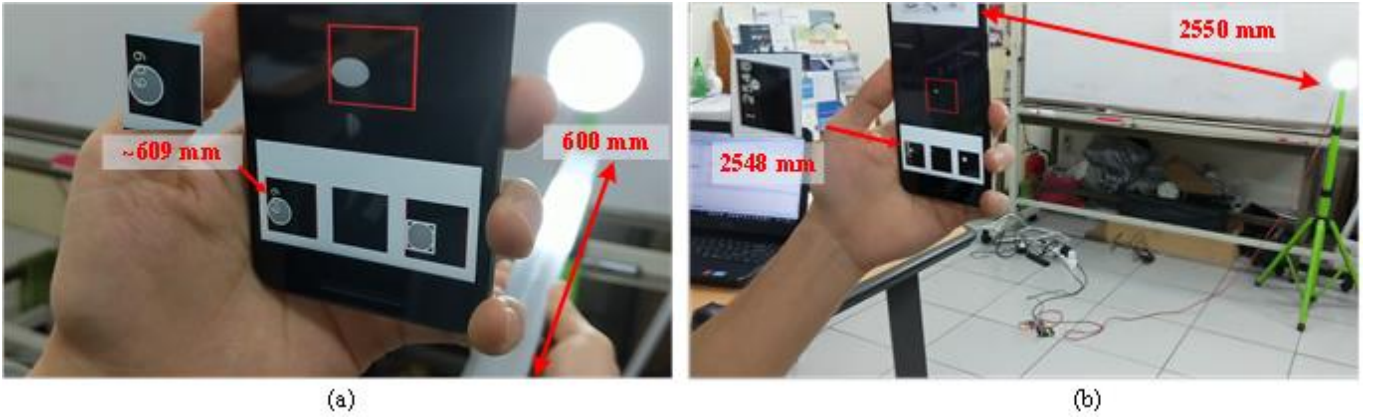

FIGURE 12: Experimental setup for distance measurement: (a) LED light at distance of 600 mm and (b) LED light at distance of 2550 mm from the camera.

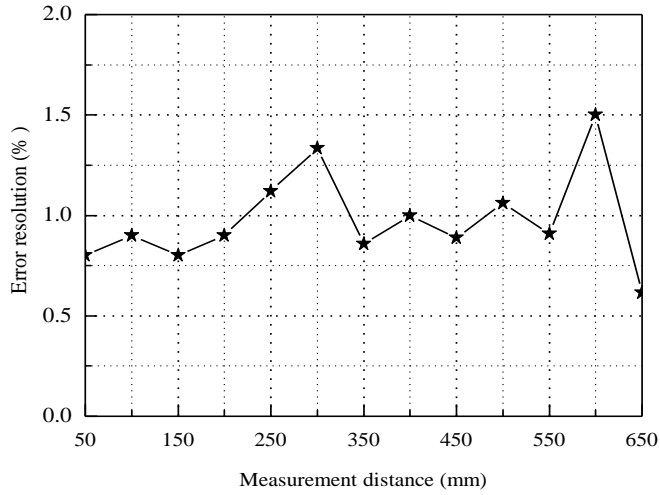

FIGURE 13: Experimental measured distance value versus error resolution.

Generally, an IS consists of a 2D pixels array of photodetector and transistors, vertical and horizontal access circuitry, and readout circuitry. Each and every pixel is accessed by the access circuitry and readout circuitry helps to read the signal value in the pixel. In dense traffic scenarios, the angular position of the FV from HV helps alleviate position measurement error. Therefore, at the middle of the IS, a plate considers as center plane as in Figure 11 which vertically separate the IS into two. With respect to this plane both angular displacement $\theta_{FV_k}$ of FV, and horizontal displacement $Hd_{IS\_FV_k}$ on the IS for corresponding FV can be measured. Here, $k^{th}(k=1,2,...,\mathbb{N})$ is the number of receive FV-ID by the HV's camera. In Figure 11, different image colors on the IS distinguish one FV-ID from other FV-IDs. The angular displacement $\theta_{FV_k}$ of FV is always zero when the FV locates at center plane. Otherwise, numerical value of angular displacement helps to mitigate some challenges from FV positioning e.g., depth estimation, lane changing information, and position estimation error mitigation for left-right side of the road. The calculated horizontal displacement $Hd_{IS\_FV_k}$ on the IS for corresponding FV is the function of FV's taillight image area $n_{IS\_FV}$ and depends on angular displacement $\theta_{FV_k}$ of FV as follows

$$Hd_{IS\_FV_k}(n_{IS\_FV} \geq \rho^2) : \left\{\theta_{FV_k}\right\} \tag{21}$$

FV's position can be determined by comparing with the position $P_{HV}$ of HV; taillight image area $n_{IS\_FV}$ of FV, horizontal displacement $Hd_{IS\_FV_k}$ on the IS for corresponding FV, and the speed of FV i.e., $\Delta V_{FV}$. Over time, these parameters will change, and consequently, the position of the FV will change. If $t$ is the initial time, then possible position of FV at $\Delta t$ is as follows

$$P_{FV}(t+\Delta t) : \left\{P_{HV}(t+\Delta t);\ n_{IS\_FV} \geq \rho^2;\ Hd_{IS\_FV_k};\ \Delta V_{FV}\right\} \tag{22}$$

TABLE 1: Transmitter and receiver parameters for simulation results.

| Parameter | Value |
|---|---|
| **Parameters for transmitter** | |
| Size of LED panel | 10x10 cm$^2$ |
| Modulation Method | S2-PSK |
| Encoding Method | Manchester coding |
| Data rate | 15 bps |
| **Parameters for receiver** | |
| Detection distance | < 30~200 m |
| Horizontal FOV | 90° - 120° |
| Image Resolution | 1 - 10 megapixels |
| Sensor physical size | 36 × 24 (mm$^2$) |
| Frame rate | 30 fps |
| Focal length | 16 - 25 mm |
| Pixel size | 2.5 - 4 μm |
| Lens aperture | 4 |
| Exposure time | 1/2000 to 1/15 (sec) |
| Height of SL | 7 m |
| Inter-distance between SL | 25 m |
| Lane width | 10 m |
| Vehicle speed | 0 - 110 km/h |

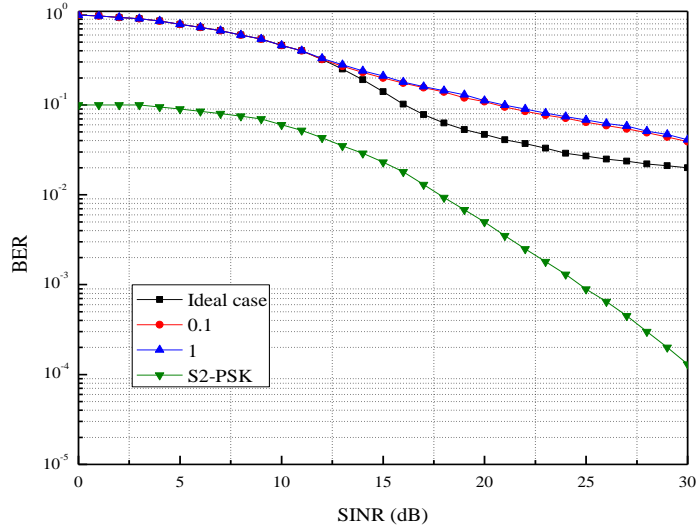

FIGURE 14: SINR versus BER.

## 3. Experimental Distance Measurement

Distance measurement using a camera is one of the important steps in the proposed scheme. Figure 12 shows the experimental setup and distance measurement procedure performed using our existing facilities under an ambient light environment. A circular LED light was used to transmit the signal. A smartphone camera was used as the receiver. With movement of smartphone, the observed distance changed. Figure 13 shows the results of experimental distance measurement. The result shows the percentage error in measurement with respect to the actual distance. The error resolution seems to remain within 1% for most distance measurements. Although the experiment could not be performed in a real vehicle environment owing to lack of facilities, this distance measurement experiment validated the feasibility of the proposed scheme.

## 4. Simulation Results

Several factors and environmental impacts must be considered for achieving localization accuracy. We considered a smooth surface to ignore the turbulence caused by vehicle movements and the impacts of other bad weather conditions (e.g., fog, snow, and rain) for generating the simulation results. The effect of a single parameter on vehicle localization accuracy was considered, whereas the other parameters were maintained constant. Table 1 lists the transmitter parameters for transmitting and summarizes the specifications of the receiver (i.e., camera) and the optical channel environment.

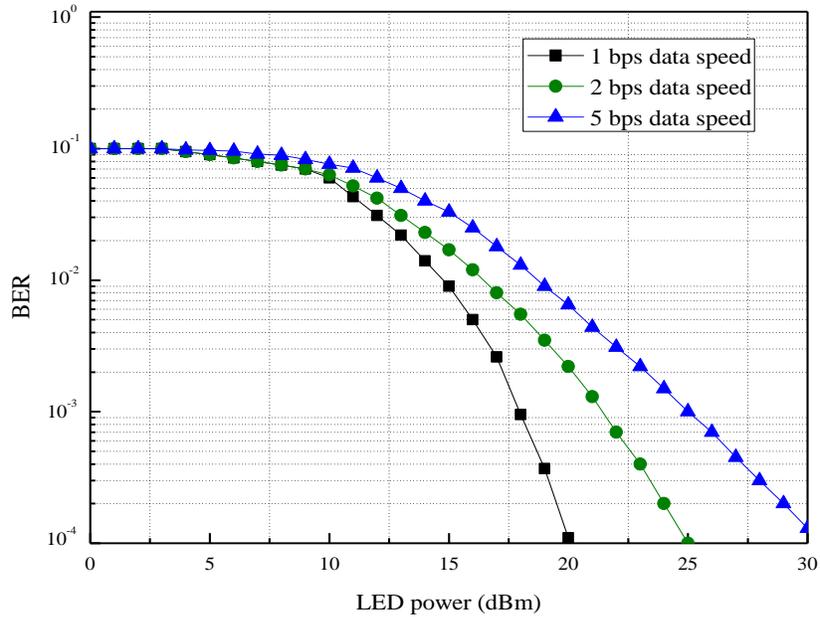

FIGURE 15: LED power versus BER.

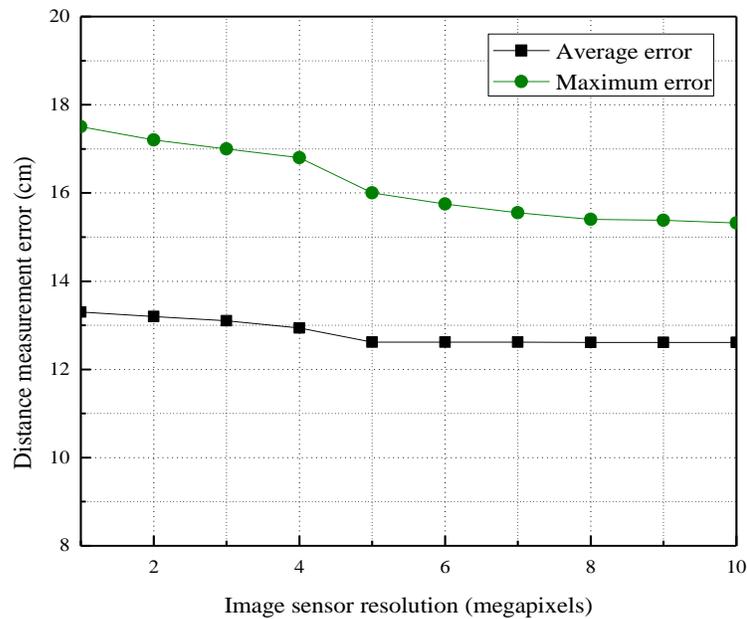

FIGURE 16: Image sensor resolution of camera with respect to distance measurement error.

A low pass filter like Gaussian filter is used to estimate the BER performance of the OCC system with respect to the SINR as a blurring filter for image processing. In this case the variance $\sigma_c^2\,(=0.5)$ for channel filtering is considered zero in ideal state. The curves for the case of estimated valiances $\sigma_c = 0.1, 1.0$ of the Gaussian filter are plotted in Figure 14 to evaluate the influence of estimation error of channel filter. In this regard, S2-PSK modulation technique based OCC system shows better BER performance with respect to SINR.

Data rate is depending on the camera frame rate. It is possible to detect one-bit data from one camera frame. For instant, 30-bit data can be received from the camera which frame per second (fps) is 30. In S2-PSK, the Manchester coding is used for data encoding. Therefore, half of total bits per second (bps) generates from 30 fps camera i.e., 15 bps. Figure 15 shows, BER performance of camera receiver with varying data speed. For simulation result is formulated for 1 bps, 2 bps, 5 bps data speed and required LED power is increased accordingly.

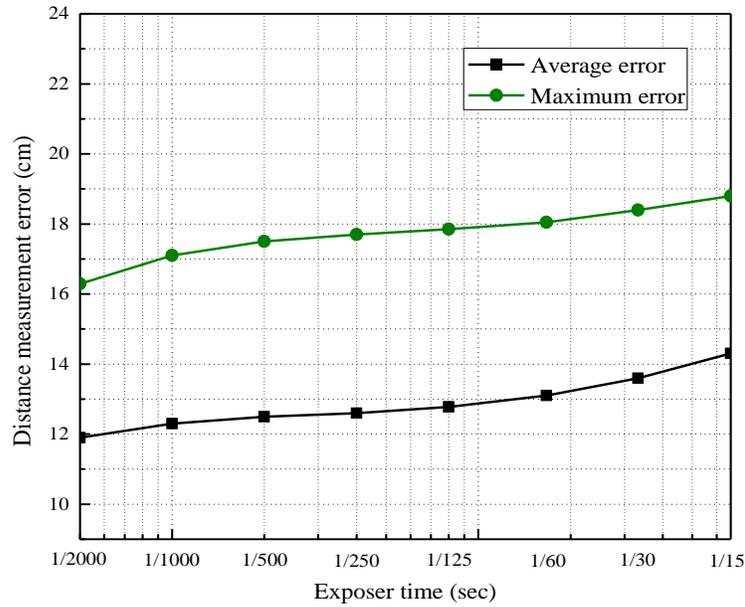

FIGURE 17: Camera exposure time with respect to distance measurement error.

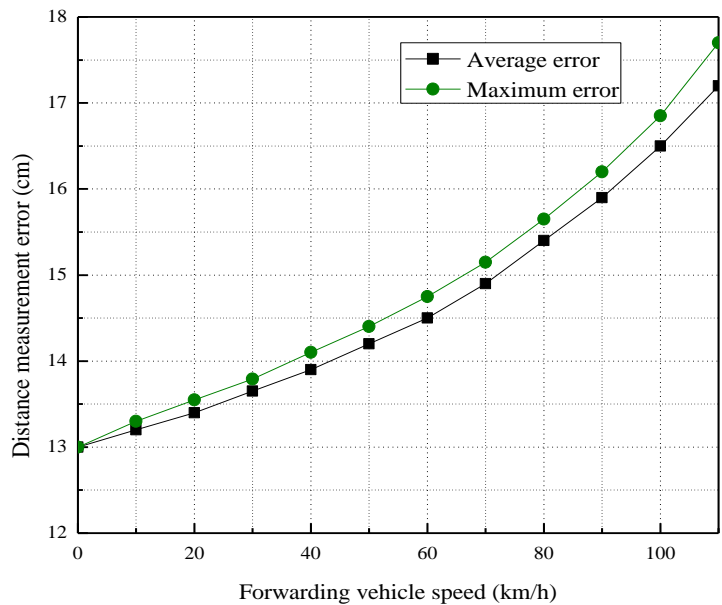

FIGURE 18: Varying speed of FV with respect to distance measurement error when its position shift maintains at zero value.

The distance error occurs when there is a discrepancy between actual and measure value of distance. Systematic error caused by environmental facts, surveillance approaches, and tool leads to this mismeasurement in such dynamic vehicular environment and need to be minimized to achieve better positioning accuracy. Average error takes from series of repeated measurements whereas maximum error generates from single measurement. The association of distance error with different camera parameters in the point of average and maximum error helps to improve the performance of distance measurement approach. IS resolution is an important camera parameter which defines by the number of total pixels; has an impact on distance calculation. It is possible to calculate the area of LED array more precisely if the camera resolution is higher. Higher resolution provides the detail about the detected LEDs to measure their area on the IS. Lower resolution will case more error in distance measurement. In Figure 16, at 1 megapixel, both maximum and average distance calculation error is higher i.e., 17.5 cm and 13.3 cm, respectively. From 5 to 10 megapixels, maximum distance measurement error is varying linearly whereas average error is fixed.

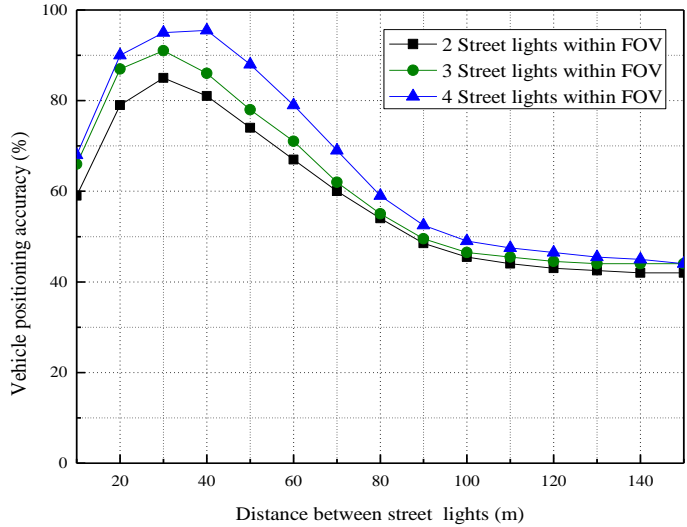

FIGURE 19: Recording the effect of distance between street lights on the measurement accuracy of HV's position.

In our proposed scheme, the camera should receive signal from LEDs at very high speed moving scenario. During this dynamic scenarios, the IS should completely expose under the illumination with every detail of the targeted LEDs i.e., street-lights, taillight of forwarding vehicles. The exposer time (or shutter speed) of the camera will ensure a period when the amount of light will be exposed on the IS. In the high speed vehicular case, large exposer time of camera will cause a blurred image and short exposer time will allow to capture detail flashes of light from a target object. Due to the dependency on the receive image quality at IS, exposer time has an impact on evaluating the performance of distance calculation. Both average and maximum distance measurement error shows equivalent evolvements with the exposer time of camera IS in Figure 17. Localization accuracy is better at lower exposer time (i.e., 1/2000). Whereas, when the exposer time is 1/15 in a second, the distance measurement error is maximum (i.e., 18.8 cm) at maximum error case.

In the mobile environment, speed and position shift; these are two important factors cause effect on vehicle distance measurement. We are considering zero shifting of FV with respect to HV to simulate the effect of speed of FV on distance calculation error. The speed of FV is varying from 0 to 110 km/h within 200 m distance whereas the speed of HV consider constant i.e., 30 km/h during simulation period which is plotted in Figure 18. At the very beginning, therefore distance measurement error will occur due to the speed of HV with respect to the FV. With the increase of the FV's speed, both average and maximum distance measurement error are increased gradually up-to 110 km/h speed. This distance measurement error occurs due to the execution time required for position calculation by the HV.

At constant vehicular speed of 50 km/h was employed considering for position accuracy measurement simulation wherein distance between SLs varied from 10 to 150 m. From Figure 19, it can be seen that as the internal distance between the SLs increases, the accuracy decreases. Moreover, at the beginning, the simulation results show that the distance measurement accuracy is relatively lower owing to the speed of the HV and data extraction from the SLs. At 50 km/h, the distance between two conjugate SLs is very small, i.e., 10 m. Therefore, within a very short period, the number of SLs crossed is greater compared to the case wherein the distance between two SLs is 40 m. At the highest point of the graph, it simplifies that the distance between SLs and receiving SL-IDs well execute to get better distance measurement accuracy i.e., near about 90%. In addition, the number of SLs influence the performance calculation. Performance improves as the number of SLs increase. This ensures the possibility of obtaining a greater number of SL-IDs at the same time and calculating the position of the HV more precisely. Furthermore, as the distance between SLs increases, the chance of comparing the location information of the SLs for accurate HV positioning is minimized. As a result, the slope of distance measurement accuracy moves downward. The variation of the lines maintains a constant margin of up to 150 m because SL-IDs have not been obtained yet.

## 5. Conclusions

A vehicle localization technique in an outdoor environment is proposed herein. The technique employs photogrammetry, which is a novel idea for localization. Implanted OCC with photogrammetry improved vehicle localization performance. The proposed technique was used to measure the distance between HVs and FVs by calculating the image area on the IS. Beforehand, the HV receives FV-IDs from each FV and uses OCC to decode these IDs. The HV's current location information helps mitigate the possibilities of relative position shifts among the HV and FV. The SL communicates with the HV in the same way as the FVs. Location information of the HV is accumulated by comparing the location of SLs with the HV's OBD II system. Experimental distance measurement confirmed the feasibility of the proposed scheme. Overall distance measurement errors were within 12–20 cm, wherein a change in one parameter was considered. The sizes of the tail LEDs of FVs are different; recognition of such LEDs is out of the scope of this study. A deep learning-based algorithm will be required to boost the performance of this single camera to overcome all challenges related to vehicle detection and localization.

## Conflicts of Interest

The authors declare no conflicts of interest.

## Acknowledgement

This research was supported by the MSIT (Ministry of Science and ICT), Korea, under the Global IT Talent support program (IITP-2017-0-01806) supervised by the IITP (Institute for Information and Communication Technology Promotion).